\documentclass[aps,prl,twocolumn,showpacs]{revtex4}
\usepackage{amsmath}
\usepackage{graphicx,epsfig,psfrag}
\usepackage{amssymb}
\usepackage{ulem}
\usepackage{hyperref}

\newcommand{\br}{\mathbf{r}}
\newcommand{\bk}{\mathbf{k}}
\newcommand{\etal}{, \text{et. al.}, }
\newcommand{\CO}{(color online).\;}

\begin{document}
\title{Equilibration rates and negative absolute temperatures for
  ultracold atoms in optical lattices}
\author{Akos Rapp}
\author{Stephan Mandt}
\author{Achim Rosch}
\affiliation{Institute for Theoretical Physics, University of Cologne, 50937
Cologne, Germany}
\date{\today}
\begin{abstract}
  As highly tunable interacting systems, cold atoms in
  optical lattices are ideal to realize and observe {\it negative}
  absolute temperatures, {\it T}  \textless { 0}. We show theoretically that by
  reversing the confining potential, stable superfluid condensates at
  finite momentum and {\it T}  \textless { 0} can be created
  with low entropy production for attractive bosons. They may serve as
  ``smoking gun'' signatures of equilibrated {\it T}  \textless  { 0}. For fermions, we analyze the time scales
  needed to equilibrate to {\it T}  \textless  { 0}. For moderate interactions, the equilibration time is proportional to
  the square of the radius of the cloud and grows with increasing
  interaction strengths as atoms and energy are transported by
  diffusive processes.
\end{abstract}

\maketitle

The concept of temperature is central in thermodynamics. For most
systems, where the energy $E$ has no upper bound, only positive
temperatures, $T > 0$, are allowed in equilibrium. However, in any
system with an upper bound in energy, $T < 0$ are possible.
In this case, states with higher energy are occupied more likely than
states with lower $E$. Negative $T$ and even phase transitions at $T<0$ have been realized in
nuclear spin systems \cite{purcell,hakonen,oja}:
antiferromagnetically coupled nuclear spins show ferromagnetic order
at $T<0$ where high-energy states are dominantly populated. Such an ``inverted'' population is also the basis of most lasers.

\begin{figure}
(a)\ \ \ \  \includegraphics[width=0.7 \linewidth,clip]{entropyEnergy.eps}\\[-0.1cm]
(b) \ \ \ \  \includegraphics[width=0.7 \linewidth,clip]{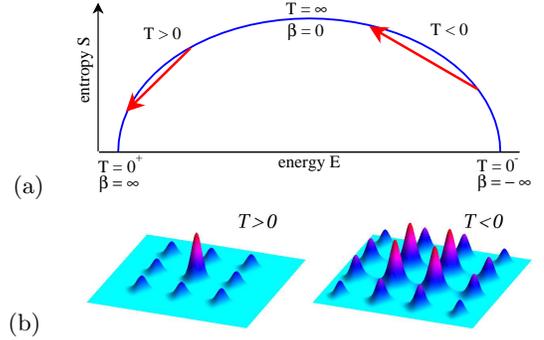}
\caption{\CO (A) Entropy as a function of energy (schematically) for a
  system with an upper and lower bound of energy. For high energies,
  $1/T=\partial S/\partial E$ is negative. Arrows: for two reservoirs
  with $T>0$ and $T<0$, respectively, one can remove reversibly energy
  from {\it both} reservoirs by reducing the entropy for $T>0$ and
  increasing it by the same amount for $T<0$. (B) Schematic
  time-of-flight image of expanding bosons. For $T>0$, repulsive
  bosons condense at momentum $0$ (with higher order peaks from Bragg
  reflections).  For $T<0$, condensates of attractive bosons form in
  the maxima of the kinetic energy for momenta $(\pm \pi/a,\pm
  \pi/a,\pm \pi/a)$.  }
\label{fig1}
\end{figure}

While negative temperatures lead to many non\-in\-tu\-itive results, all
laws of thermodynamics can equally be applied
\cite{ramsey}. Figure \ref{fig1}(a) shows schematically the entropy $S$ as a
function of energy for a system with a maximal and minimal energy. As
$1/T = \partial S/\partial E$, negative $T$
arises whenever the entropy decreases as a function of energy.
One consequence is that a Carnot engine, which operates between
two reservoirs with temperatures $T_1<0$ and $T_2>0$, has an efficiency $\eta$ larger than 1: $\eta = W/Q_1 = 1- T_2/T_1 > 1$,
where $W$ is the work done and $Q_1$ is the heat extracted from the (hotter) reservoir.
Usually, $\eta<1$ as the entropy $\Delta S$ extracted from the hot reservoir has to be dumped
into the second reservoir and the corresponding heat $T \Delta S$ is lost.
If the first reservoir has, however, negative $T$, its
energy is {\it lowered} when the entropy increases, see Fig.~\ref{fig1}(a): heat
can be extracted from both systems simultaneously and therefore $\eta>1$.

With quantum-optical methods it is possible to ma\-ni\-pulate atoms in an
unprecedented way. Atoms in optical lattices allows us not only to realize Hubbard models \cite{jaksch} for
bosons \cite{MI-SF-paper} and fermions \cite{Esslinger, Bloch}, but also to control
all parameters  with high precision. In the following, we will first show how simple manipulations of such systems can be used to create
and detect unambiguously equilibrated quantum states at negative $T$,
following partially a proposal of Mosk \cite{mosk}, and
then we discuss in the case of fermions the time scales needed for equilibration.

Atoms in sufficiently deep optical lattices can be described
by Hubbard models \cite{jaksch} given by
\begin{eqnarray}
H_f & =&-J \sum_{<ij> \sigma} ( f^{+}_{i\sigma} f^{\phantom{+}}_{j\sigma} + {\rm h.c.} )
     + U \sum_i n_{i\uparrow} n_{i\downarrow} \nonumber \\
 &&\qquad + V_0(t) \sum_{i \sigma} \br_i^2 n_{i\sigma},
	 \label{eq:fermiHubbard}   \\
H_b &=&  -J\! \sum_{<ij>} \!(b^{+}_{i} b^{\phantom{+}}_{j} + {\rm h.c.}) + \frac{U(t)}{2} \sum_i n_{i} (n_{i}-1) \nonumber \\
 &&\qquad + V_0(t) \sum_{i} \br_i^2 n_{i}\;, \label{eq:bosonHubbard}
\end{eqnarray}
for fermions and bosons, respectively. $U$ is the local interaction, $J$ the tunneling rate and $\sigma = \uparrow, \downarrow$ the hyperfine index for the fermions. $V_0(t)$ represents a time-dependent parabolic potential with $n_{i\sigma} = f^{+}_{i\sigma} f^{\phantom{+}}_{i\sigma}$, $n_{i} = b^{+}_{i} b^{\phantom{+}}_{i}$. In Eqs.~(\ref{eq:fermiHubbard}-\ref{eq:bosonHubbard}) higher bands are omitted. As discussed in detail by Mosk\cite{mosk}, typical tunneling rates into such bands are exponentially suppressed and negligible.

Negative $T$ in equilibrium is only possible for Hamiltonians bounded
from above. Thus, for the
models~(\ref{eq:fermiHubbard}-\ref{eq:bosonHubbard}), $V_0<0$ and for
bosons also $U<0$ is required. Nevertheless, ultracold atoms have to be prepared with some $V_0 > 0$ initially, and we shall
discuss how $T<0$ can be reached from such conditions (preparation of
high-energy states in spin-systems is discussed in Ref.~\cite{demler}).

To get some intuition on negative $T$, note that the equilibrium
density matrix, $e^{-H/k_B T}$, for a Hamiltonian $H$ at $T < 0$ is
identical to the density matrix for reversed temperature $\tilde T =
-T$ and a Hamiltonian $\tilde H=-H$.  For the
models~(\ref{eq:fermiHubbard}-\ref{eq:bosonHubbard}) therefore
considering $T<0$ is equivalent to $\tilde T > 0$ with parameters
$-V_0$, $-U$, and most importantly, $-J$.  As for a cubic lattice with
lattice constant $a$ the
sign of $J$ can be absorbed into a shift of all momenta, $k \to k + Q$
with $Q=(\pi/a,\pi/a,\pi/a)$ [using \cite{Boltzmann-eq} that $-J
\cos(k a)=J \cos(k a+ \pi)$], the phase diagram of the negative-$U$
Hubbard models for $T<0$ are identical to that of the positive-$U$
Hubbard models if the momenta are shifted in all
observables. Most dramatically, bosons will therefore
condense at momenta $(\pm \pi/a,\pm \pi/a,\pm \pi/a)$, i.e., in the
maxima of the band structure, for $U<0, T<0$ [ see
Fig.~\ref{fig1}(b)]. Note that $T<0$ bosonic condensates are stable for
attractive and unstable for repulsive $U$. For fermions superfluidity
can also be reached for $T<0, U>0$ but in this case the fermionic
pairs condense at zero momentum  as $\sum_k f^+_{k \uparrow} f^+_{-k
  \downarrow}=\sum_k f^+_{k+Q \uparrow} f^+_{-k-Q
  \downarrow}$. We remark that it has been shown \cite{etapairing}
that for $U \gg J$ one can nevertheless induce a condensate at
momentum $Q$ even for fermions in a 
different nonequilibrium situation. To summarize, the
momentum distribution of a Bose-Einstein condensate is probably
the best way to detect $T<0$ due to the {\it qualitative} difference
to the $T > 0$ system, and we will quantitatively estimate under what conditions such a state can be reached.

In principle, it is possible to reach $T<0$ {\it without} any entropy production by first {\it
adiabatically} decreasing $J$ until $J=0$, then
switching suddenly $U \to -U$ and $V_0 \to -V_0$, followed by an {\it adiabatic} increase of $J$ until the
desired value is reached. In practice this is not a good choice as the timescales for equilibration diverge for $J \to 0$. Therefore one needs a faster scheme where $J$ remains finite but as little entropy as possible is produced. As a reversal of $V_0, U$ and $T$ is formally equivalent to a sudden quench $J \to -J$ without inverting $T$ (see above), the basic idea is to start from an initial state where $J$ is finite but the kinetic energy is very small, i.e., a Mott or a band insulator.

For bosons we propose the following. The system is (I) prepared
in a Mott-insulating state. (II) $J$ is switched off
{\it suddenly} by increasing the intensity of the optical
lattice. This freezes all density-density correlation functions. One waits (III) for a time $t_{\rm w}$ during which the potential and the interaction are reversed slowly, $V_0 \to -V_0$ and $U \to - U$.
(IV) $J$ is switched  suddenly back to its initial value and (V) the system equilibrates. Finally, (VI) the negative trapping potential and/or the interaction strength $U$ is weakened {\it adiabatically}
 with the goal to reach a superfluid condensate at $T<0$.
 For fermions, parts of this scheme have been implemented in Ref.~\cite{Boltzmann-eq}, where, however, $V_0$ was not reversed but set to $0$.

The waiting time $t_{\rm w}$ allows us to reverse $V_0$ and $U$ slowly
(manipulation of $U$ requires ``slow'' changes of magnetic fields in
some experimental setups), and more importantly, it leads to a
complete dephasing of the kinetic energy, $E_{\rm kin}$, as has been
argued in Ref.~\cite{Boltzmann-eq}: During $t_W$ each site in the
lattice collects a different phase due to $V_0$. Effectively,  therefore $E_{\rm kin}\propto \sum_{\langle
 ij\rangle} \langle b^\dagger_i b_j\rangle$ averages to zero for $t_W \gg 1/\Delta V$, 
where $\Delta V$ is the potential difference between neighboring
sites. Note that for a Mott insulator $E_{\rm kin}$ vanishes in the
center of the trap but is sizable in the outer parts of the cloud
where $\Delta V$ is large. Estimates using the parameters of Ref.~\cite{MI-SF-paper} show that waiting times $t_{\rm w}
\gtrsim 10$\,ms are sufficient for a complete dephasing.

For a quantitative estimate of the entropy generated during this sequence, one first
needs to know the total energy $E_{\rm tot}$
after step (V), i.e., before the final equilibration stage. As
the kinetic energy is zero after dephasing and the initial density distributions are fully preserved, we obtain $E_{\rm tot}= -(E_{\rm int}+E_{\rm
pot})$, where $E_{\rm int}$ and $E_{\rm pot}$ are the {\it initial} interaction and potential energies,
respectively. We obtain these variationally using the Gutzwiller wave function for
bosons \cite{boson-Gutzwiller} assuming a low-$T$ initial state for a given strength of the confining
potential characterized by the
dimensionless compression $V_0 N^{2/3}/6J$, where $N$ is the number of atoms. This combination appears
naturally \cite{Bloch} as the thermodynamic limit in a trap is given
by $N \to \infty$ with $V_0 N^{2/3}=const.$

Energy conservation allows us to determine the state after equilibration. The corresponding temperature will be negative and large. We therefore used a high-temperature expansion \cite{highTexp} to obtain the thermodynamic potential in the presence of the trap. We found that an expansion up to 4$^{th}$ order in $\beta J$ gives accurate results. For simplicity, we locally approximated the system by a homogeneous one.
Corrections to this so-called local-density approximation \cite{LDA} vanish with $1/N^{1/3}$ and are therefore
tiny for typical atom numbers, $N \sim 10^5$. Finally, the (negative) temperature was determined by setting
$\langle H_b \rangle=E_{\rm tot}$ and then the entropy was computed from the high-$T$ expansion.

\begin{figure}
\includegraphics[width=0.9 \linewidth,clip]{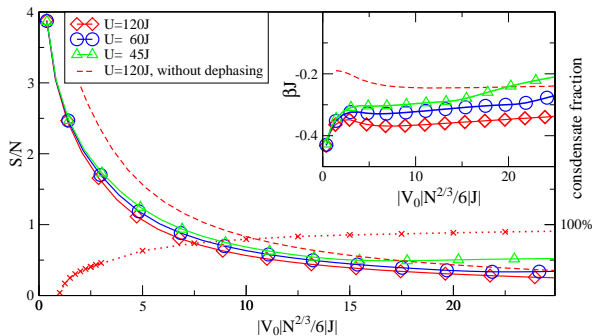}
\caption{\CO Entropy per boson as a function of the initial
  confinement $V_0$, for different initial interactions $U$ after following the steps (I)-(V), see text. Dotted line: Condensate fraction
at momenta $(\pm \pi/a,\pm \pi/a,\pm \pi/a)$ for $U=120\,J$ after a slow decrease of $|U|$ during step (VI). A slightly higher condensate fraction is obtained, if in step (VI) $|V_0|$ is reduced (not shown).
Inset:  inverse temperatures. Dashed lines: results for
$U=120 \,J$ using the protocol suggested by Mosk \cite{mosk}, i.e., without dephasing of the kinetic energy.
}\label{fig:TDfit}
\end{figure}

In Fig.~\ref{fig:TDfit} the entropy per boson is shown as a function of the initial compression $V_0 N^{2/3}$
of the cloud for several values of $U$. We also calculated the result without dephasing (dashed lines), obtained in
the limit $t_{\rm W} \to 0$, corresponding to the scheme originally proposed by Mosk~\cite{mosk}. Entropies and $|T|$ are about $40\%$ higher in this case.

The entropy per boson drops for increasing $| V_0 |$ as larger and larger fractions of the cloud become Mott
insulating and thus insensitive to sign changes of $J$. Most importantly, even for moderate compressions and
moderate values of $U$ the entropy per boson is well below the critical entropy of an ideal Bose gas in a
$d=3$ harmonic trap, $S/N < s_0 = 3.6 \,k_B$. This implies that for realistic parameters, one can expect the
formation of the negative-$T$ Bose-Einstein condensate after the adiabatic expansion [step (VI)] characterized by the momentum
distribution of Fig.~\ref{fig1}B and by a large condensate fraction. Figure \ref{fig:TDfit} shows the condensate
fraction assuming that in step (VI) $|U|$ has been reduced adiabatically to $0$ for fixed $V_0<0$.

We will now consider spin-$1/2$ fermions focusing on dimensions $d=2$ to
simplify numerics. With fermions one can reach $T<0$ using a
time-independent $U$ starting from a band insulating, rather than a
Mott-insulating initial state. We will focus our discussion on the
time scales needed to reach $T<0$. To determine the dynamics, we use a
numerical solution of the Boltzmann equation in
relaxation time approximation, \cite{Boltzmann-eq}
\begin{equation}
    \partial_t f + \mathbf{v}_\bk \cdot \nabla_\br f + \mathbf{F} \cdot \nabla_\bk f = -\frac{1}{\tau(n,e)}( f - f_0(n,e)) \;,
\label{eq:Boltzmann}
\end{equation}
where $f(\br,\bk,t)$ is the occupation probability in phase space and
the force term $\mathbf{F} = - \nabla_\br (V_0 \br^2) - U \nabla_\br
n(\br)$ contains the external potential and interaction corrections on
the Hartree level. $n=n_{\uparrow}=n_{\downarrow}=n(\br,t)$ and $e=e(\br,t)$
 are the local particle densities per spin and energy density, respectively, and $f_0(n,e)$ is a Fermi
function chosen such that energy and particle number are conserved.
In Ref.~\cite{Boltzmann-eq} we have determined the scattering rate
$1/\tau(n,e)$ to reproduce the $e$- and $n$-dependent diffusion
constant of the Hubbard model to order $U^2$ (obtained from an
independent calculation). Most importantly, ${1/\tau(n,e)}$ vanishes
for low densities, ${1/\tau(n,e)}\propto n$, in the tails of the cloud
where scattering is rare. In Ref.~\cite{Boltzmann-eq} the numerics was also
compared to experiments.

\begin{figure}[ht]
    \centering
    \includegraphics[width=0.9 \linewidth,clip]{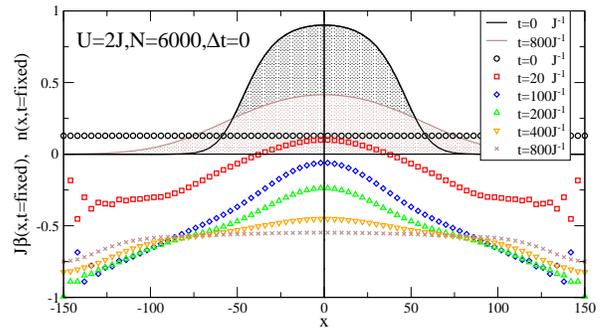}
    \caption{\CO Cuts through the density $n$ (lines) and inverse
          temperature $\beta=1/k_B T(\br,t)$ (symbols) for quenches $V_{0i} \to -0.05 V_{0i}$ at $U=2J$, $V_0 N=60 J$,
          initial entropy $S/N=1.2 k_B$ and $N=6000$ particles per spin.}
\label{fig:RT0}
\end{figure}

We first consider an instantaneous quench, $V_{0i} \to V_{0f}=-0.05
V_{0i}$ for  $U =2J$ and parameters which reflect approximately the experimental conditions of Ref.~\cite{Boltzmann-eq}. Temperature and density profiles (and parameters) are displayed in Fig.~\ref{fig:RT0}.
Here $T(\br,t)$ is defined as the temperature of a homogeneous reference
ensemble in equilibrium with the same energy and particle density.

Because of energy conservation and the upper bound on the kinetic energy,
the atomic cloud cannot expand to arbitrary size but equilibrates for $t \to \infty$. The initial $T>0$ becomes rapidly negative and slowly obtains a homogeneous value (which can be calculated from energy conservation in an independent way). It takes longer to equilibrate in the tails of the cloud, where scattering rates are small and strong Bloch oscillations occur.

\begin{figure}[ht]
    \centering
    \includegraphics[width=0.9 \linewidth,clip]{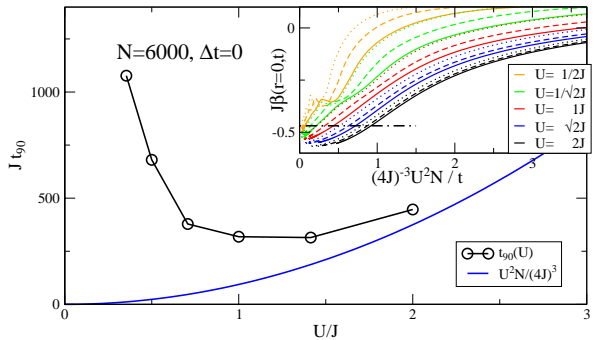}
    \caption{\CO Time required to reach $\beta_0 = 0.9 \times
          \beta_f(U\to 0)$ for $N=6000$. Inset:  $\beta(\br = 0,t)$ in the center of the trap as a function of
          rescaled $1/t$ for various $U$ and system sizes (solid: $N=3000$, dashed: $N=6000$, dotted: $N=12000$). Dash-dotted line:  $J\,\beta_0 = -0.469$.  }
\label{fig:optimU}
\end{figure}

Two time scales determine the relaxation mainly. First,
$\tau(n,e)$ determines locally the equilibration rate according to
Eq. (\ref{eq:Boltzmann}) but local interactions do not change the
local energy and particle density. Therefore a second time scale
describes how long it takes to redistribute particles and energy
across the cloud. For not too weak interactions, see below, this
transport will be diffusive and not convective as momentum is {\it not} conserved and can be efficiently transferred
to the optical lattice by umklapp scattering for the relatively high $|T|$ considered here. The corresponding diffusion time is estimated as
\begin{eqnarray}
\tau_D \sim \frac{r^2}{D} \sim \frac{U^2 N}{J^3}
\end{eqnarray}
where $r$ is the radius of the cloud, $D\sim v^2 \tau$ the diffusion
constant, $v \sim J a$ the typical velocity (for $T\sim J$), $1/\tau
\sim n U^2/J$ and
$n (r/a)^2\sim N$. The inset of Fig.~\ref{fig:optimU} displays $1/T$ in
the center of the cloud as a function of $\tau_D/t$. It shows that $\tau_D$ becomes the relevant
time scale for large $N$ and not too small $U$. Furthermore, for $t \ll
\tau_D$, where the edges of the cloud do not yet play a role,  there
is a regime,
where  $\beta \sim 1/t\,$ is observed as expected for (energy) diffusion in $d=2$.
Figure \ref{fig:optimU} also shows the time $t_{90}(U)$ needed
to reach $J \beta_0=-0.469$ ($90\%$ of the inverse temperature for $t\to \infty, U\to 0$). For larger $U$ one gets $t_{90}\sim \tau_D \propto U^2$ but for small $U$, $t_{90}$ diverges due to the divergence of the local relaxation time, $\tau \propto 1/U^2$. In all cases, relaxation is not very fast, $t_{90}>300/J$. The fastest relaxation to equilibrium occurs for relatively weak interactions when $U$ is a fraction of the full bandwidth $8 J$.

\begin{figure}[ht]
    \centering
    \includegraphics[width=0.9 \linewidth,clip]{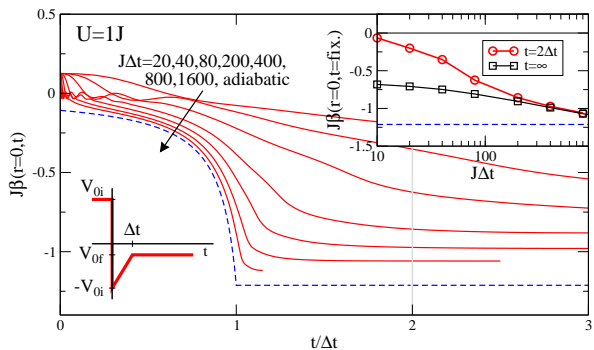}
    \caption{\CO The inverse temperature $\beta(\br=0,t)$ in the
          center of the cloud as a
          function of $t/\Delta t$ for different opening times $\Delta
          t$, $U=J$ and a $V_0(t)$ shown schematically in the left
          inset ($V_{0f}=-0.05 V_{0i}$, $V_{0i} N=60\,J$, $N=6000$). Right inset: Inverse
        temperatures  $\beta(\br=0,t=\infty)$ and  $\beta(\br=0,t=2
        \Delta t)$ as a function of $\Delta t$.}
\label{fig:RTchange}
\end{figure}


In order to reach $T<0$ with small $|T|$, it is useful to decrease $|V_0|$ slowly to reduce entropy generation and to see whether adiabatic conditions can be realized. We therefore use the protocol shown in the inset of Fig.~\ref{fig:RTchange}: after a sudden quench at $t=0$, $V_{0i} \to -V_{0i}$, $-V_0$ is reduced linearly, $V_0(t)=-V_{0i}+(V_{0f}+V_{0i}) t/\Delta t $ for $t<\Delta t$ and  $V_0(t)=V_{0f}<0$ for $t>\Delta t$. As shown in Fig.~\ref{fig:RTchange}, upon increasing $\Delta t$, considerably lower values of $\beta<0$ can be obtained and one approaches the adiabatic limit. Only due to the high entropy assumed for the initial state ($S/N=1.2\,k_B$, implying $J\, \beta \approx 1.47$ for $V_{0i} \to |V_{0f}|$ adiabatically), $|T|$ remains relatively high even for $\Delta t \to \infty$ where $J\, \beta \approx -1.21$. The overall entropy production for $\Delta t \to \infty$ is tiny, $\Delta S/N\approx 0.12 \,k_B$, [c.f. the bosonic case, Fig.~\ref{fig:TDfit}] as for the large initial $T$, kinetic energies were small. Note that
even for $\Delta t=1600/J$ deviations from the adiabatic behavior are
considerable, which shows how difficult it is to reach truly
adiabatic conditions. Nevertheless, it is possible to
reach  $T\approx -2 J$ within a time $200/J \sim 100\, {\rm ms}$ for typical parameters~\cite{Boltzmann-eq}.

In our opinion the observation of finite momentum superfluidity,
Fig.~\ref{fig1}B, is probably the best ``smoking gun'' signature of
$T<0$ in equilibrium. To reach it, it is, however, necessary to switch
the interaction $U\to -U$ using a Feshbach resonance for bosons. We
expect that the associated loss processes by three-particle scattering can
efficiently be reduced in an optical lattice.  An important issue are
the time scales needed for local equilibration and -- most importantly
-- redistribution of energy and particles across the system. For
fermions we find that relaxation is most efficient for relatively weak
interactions.  More generally, the long equilibration times arising
from the necessity to redistribute energy and particles should be
important for all equilibration processes and quenches in
inhomogeneous systems both for positive and negative $T$. Here we
expect that the equilibration properties of low-$|T|$ bosons differ
qualitatively from high-$|T|$ fermions due to the suppression of umklapp
scattering for bosons and due to the onset of superfluidity.

\acknowledgements We acknowledge discussions with
I. Bloch, D. Rasch, E. Demler, and U. Schneider, and financial support by the SFB 608 and SFB/TR 12 of the DFG and the Studien\-stiftung des deutschen Volkes  (S.M.).

\end{document}